\documentclass[submission,copyright,creativecommons]{eptcs}
\providecommand{\event}{HSB 2013} 
\usepackage{breakurl}             

\usepackage{amssymb}
\usepackage{amsfonts}
\usepackage{amsmath}
\usepackage{amsthm}
\usepackage{graphicx,caption,subcaption}

\usepackage{MnSymbol}

\usepackage{tikz} 
\usetikzlibrary{arrows,automata,shapes} 
\tikzstyle{every picture}+=[remember picture]
\usepgflibrary{decorations.pathmorphing} 
\usepgflibrary[decorations.pathmorphing] 
\usetikzlibrary{decorations.pathmorphing} 
\usetikzlibrary[decorations.pathmorphing] 

\title{Falsifying Oscillation Properties \\ of Parametric Biological Models}

\author{
Thao Dang
\institute{CNRS-VERIMAG, Grenoble, France}
\email{Thao.Dang@imag.fr}
\and
Tommaso Dreossi
\institute{VERIMAG, University Joseph Fourier, Grenoble, France}
\institute{Dept. of Mathematics and Computer Science\\
University of Udine, Udine, Italy}
\email{Tommaso.Dreossi@imag.fr}
}
\def\titlerunning{Falsifying Oscillation Properties of Parametric Biological Models}
\def\authorrunning{T. Dang \& T. Dreossi}

\newcommand{\SLinPred}{\lambda}
\newcommand{\LinPred}{\pi}

\newcommand{\absFct}{\alpha}
\newcommand{\abStSpace}{S}
\newcommand{\absTrans}{\leadsto}
\newcommand{\concFct}{\gamma}
\newcommand{\absst}{s}
\newcommand{\X}{\mathcal{X}}

\newcommand{\bool}{\mathbb{B}}
\newcommand{\real}{\mathbb{R}}

\def \ttrans#1 {\stackrel{#1}{\rightarrow}}
\def\real{\mathbb{R}}

\newcommand{\cfct}{\mathcal{U}}

\newcommand{\se}{\subseteq}

\newcommand{\node}{v}

\newcommand{\xgoal}{x_{goal}}
\newcommand{\xnear}{x_{near}}

\newcommand{\Inv}{\mathcal{I}}
\newcommand{\Guard}{\mathcal{G}}

\newcommand{\stspace}{\mathcal{X}}
\newcommand{\A}{\mathcal{A}}
\newcommand{\dA}{\mathcal{D}}

\newtheorem{definition}{Definition}

\begin{document}
\maketitle

\begin{abstract}
We propose an approach to falsification of oscillation properties of parametric biological models, based on the recently developed techniques for testing continuous and hybrid systems. In this approach, an oscillation property can be specified using a hybrid automaton, which is then used to guide the exploration in the state and input spaces to search for the behaviors that do not satisfy the property. We illustrate the approach with the Laub-Loomis model for spontaneous oscillations during the aggregation stage of Dictyostelium. 
\end{abstract}


\section{Introduction}\label{sec:introduction}

Understanding periodic responses in living organisms is an important problem since such oscillations are a common phenomenon in biology. To reveal possible molecular mechanisms underlying this phenomenon, mathematical models have been developed. These models require validation before they can be used to make predictions. Such validation is often based on a comparison between the model behavior and experimental data obtained by temporal measurements. One major difficulty in biological model validation is that biological models often require many parameters, and most parameter values are neither measurable nor available in literature. Since there are often many sets of parameter values that can match the data, parameter identification is based not only on the error between the model simulation output and the data, but also on model robustness with respect to parameter variation. From a modelling point of view, robust parameters allow the model to fit new data without compromising the fit to the previous data. From a biological point of view, with robust parameters the system is resilient to perturbations. 

The focus of this work is twofold. On one hand, we are interested in studying biological oscillating behaviors. On the other hand, we want to study the influence of parameters on the system behavior, that is how much the parameters can be varied without violating a given property. Typical behavioral changes include self-oscillations (that is the developments of periodical orbits from an equilibrium) and the occurence of a bifurcation. To illustrate this, we consider a dynamical system described by the following differential equations:
$$\dot{x} = f(x, k)$$
where $x \in \real^n$ is the state variables, and $k \in \real$ is a real-valued parameter. In a more general context, the dynamics of the system can be hybrid and contain more than one parameter. To characterize the impact of parameter variation, we want to know under which condition the two systems $\dot{x} = f(x, k)$ and $\dot{x} = f(x, k')$ (under two different parameter values $k$ and $k'$) are qualitatively similar, that is there exists an inversible and continuous homeomorphism that maps a trajectory of one system to a trajectory of the other. In this case, an oscillating trajectory and a steady state equilibrium of one system correspond respectively to an oscillating trajectory and a steady state equilibrium of the other. When the parameter changes and reaches a value at which the behaviors are no longer qualitatively similar, such a behavioral change is often called a bifurcation and this value is called a critical value or a value of bifurcation. This can be illustrated with a linear dynamical system when the real parts of its eigenvalues change their sign under a parameter variation. 

A set of parameter values is called {\em robust} if the system does not undergo a bifurcation under any variation of the parameter within that set. In this paper we propose to use a testing approach to analyze the robustness of biological models with respect to preservation of oscillation properties under admissible parameter variations. When applied to a model, this testing approach can be seen as systematic simulation that can check whether the model can replicate some essential behaviors observed during experiments. However, in general it can also be applied to a biological system (viewed as a black-box system). The key steps of the approach we propose are the following:
\begin{enumerate}
\item {\em Specifying the property}. A hybrid automaton $\A$ is used to describe the expected oscillating behaviors. We call $\A$ a \emph{property automaton}. This automaton also encodes the satisfaction/violation of the property and incorporates realistic variations of the parameter values. 
\item {\em Generating test cases for property falsification.} The generation of test cases from the automaton $\A$ is randomized but guided by the property, that is it favors the exploration of the trajectories leading to a violation of the property of interest.  
\end{enumerate}

We choose hybrid automata as specification formalism for two reasons. First, numerous phenomena in biology exhibit switching behaviors. Second, hybrid automata can naturally describe transitions between different qualitative behaviors, as we will show later. In the hybrid systems research, formal specification of oscillation properties of biological systems are considered in \cite{Bartocci2009,Brim2012}.

Concerning bifurcation detection, the theory of bifurcation in smooth systems is well developed. The existing methods (such as using analysis of the eigenvalues of the Jacobian matrix \cite{Iglesias2006}, Routh-Hurwitz stability criteria \cite{4434348,DelVecchio2009}, the Floquet multipliers \cite{Kuznetsov2004}) are developed for continuous systems and it is not easy to extend them to hybrid systems with discontinuities in the dynamics. Another approach (such as \cite{Donahue2009}) involves first generating the model outputs by simulation and then finding the parameters by fitting the simulation outputs to the experimental data, based on a grid over the parameter space. Our testing-based approach with a property guided search enables a quick detection without exploring a large number of parameter values. In addition, the approach has the potential to be more scalable than analytical and grid-based methods.
 
The rest of the paper is organized as follows. We first describe how to use hybrid automata to specify oscillation properties. This specification formalism can be applied to a large class of temporal properties due to the expressiveness of hybrid automata. We then show how to generate test cases from a property automaton for falsification purposes. The approach is applied to analyze the robustness of the Laub-Loomis model under parameter variation. This model has been proposed for describing the dynamical behavior of the molecular network underlying adenosine 3'5'-cyclic monophosphate (cAMP) \cite{laub1998molecular}.

\section{Using Hybrid Automata to Specify Oscillation Properties}\label{sec:temporal_hybrid_automata}

We first present a commonly used definition of hybrid automata and then show how they can be used to for oscillating property specifications.

\subsection{Hybrid Automata}\label{sec:hybrid_automata}

In the development of formal models for designing engineering systems, hybrid automata~\cite{Alur} emerged as an extension of timed automata \cite{alur94} with more general dynamics.
Unlike in a timed automaton where a clock $c$ is a continuous variable with time derivative 
equal to $1$, that is $\dot{c} = 1$, in a hybrid automaton its continuous variables $x$ can evolve
according to some differential equations, for example $\dot{x} = f(x)$.
This allows hybrid automata to capture the evolution of a wide range of physical entities.

\begin{definition}[Hybrid automaton] A {\em hybrid automaton} is a tuple $\A = (\stspace, Q, E, F, \Inv, \Guard, \Reset (q_{0}, x_{0}))$ where  \begin{itemize}
\item $\mathcal{X} \subseteq \mathbb{R}^n$ is the continuous state space;
\item $Q$ is a finite set of locations (or discrete states);  
\item $E \subseteq  Q \times Q$ is a set of discrete transitions;  
\item $F=\{ F_q ~|~  q \in Q \}$ specifies for each location a \emph{continuous vector field}. In each location $q \in Q$ the evolution of the continuous variables $x$ are governed by
a differential equation $\dot{x}(t) = f_q(x(t),u(t))$ where $u(\cdot) \in \mathcal{U}_q$ is an input function of the form $u : \mathbb{R}^+ \rightarrow U_q \subset \mathbb{R}^m$. The set $\cfct_q$ is the set admissible inputs and consists of piecewise continuous functions.  We assume that all $f_q$ are Lipschitz continuous;  
\item $\mathcal{I} =\{ \mathcal{I}_q \subseteq \mathcal{X} ~|~ q\in Q \}$ is a set of invariants. The \emph{invariant} of a location $q$ is defined as a subset $\mathcal{I}_q$ of $\mathcal{X}$. The system can evolve inside $q$ if $x \in \mathcal{I}_q$;
\item $\mathcal{G}=\{ \mathcal{G}_e ~|~ e \in E \}$ is a set of \emph{guards} specifying the conditions for switching between locations. For each discrete transition $e=(q,q')\in E$, $\mathcal{G}_e \subseteq \mathcal{I}_q$; 
\item $\mathcal{R}= \{ \mathcal{R}_e ~|~ e \in E \}$ is a set of \emph{reset} maps. Each transition $e=(q,q')\in E$ is associated with a reset map $\mathcal{R}_e: \Guard_e \to 2^{\Inv_{q'}}$ that defines how $x$ may change when the automaton $\A$ switches from $q$ to $q'$; 
\item The initial state of the automaton is denoted by $(q_{0}, x_{0})$. 
\end{itemize}
\end{definition}

A state $(q,x)$ of $\mathcal{A}$ can change in the following two ways:
\begin{enumerate}
	\item	by a \emph{continuous evolution}, where the continuous state $x$ evolves according to the dynamics $f_q$ while the location $q$ remains constant;
	\item by a $\emph{discrete evolution}$, where $x$ satisfies the guard of an outgoing transition and the system changes location by taking this transition and updating the values of
		$x$ accordingly to the associated reset map.
\end{enumerate}

It is important to note that hybrid automata allow modelling non-determinism in both continuous and discrete evolutions. This non-determinism is useful for describing disturbance from the environment or under-specified control, as well as for taking into account imprecision in modeling.

\subsection{Property Automata}\label{sec:prop_bio_systems}

We now show how to formalize some common temporal properties of particular interest for biological systems using hybrid automata. We will call them \emph{property automata}.

A dynamical system starting from a given initial state can evolve to a steady state or to an irregular behavior. The steady state may be stationary (that is, the system remains in the same state as time passes), which is also called an equilibrium. The system can also evolve to a periodic state (or a limit cycle). Stationary states and periodic states can be stable (that is, attracting neighboring trajectories), unstable (that is, repelling neighboring trajectories), or non-stable (saddle). The stationary and periodical states are important since they help determine the long-term behavior of the system. It is often of great interest, in particular for biological systems, to know how the stationary and periodic states change when the parameters of the system change. 

\begin{figure}
	\begin{center}
		\begin{tikzpicture}[shorten >=1pt,node distance=2cm, auto]
	
		\node[state] at (5,5) (q_1) {\Large $\substack{\dot{x} = f(x,k) \\ \dot{k} = u \\ \dot{c} = 1 \\ \dot{p} = 0 \\ \dot{x_p} = 0}$};
		\node[state,initial] at (0,0) (q_2) {\Large$\substack{\dot{x} = f(x,k) \\ \dot{k} = u \\ \dot{c} = 1 \\ \dot{p} = 0 \\ \dot{x_p} = 0}$};
		\node[state] at (5,0) (q_3) {\Large$\substack{\dot{x} = f(x,k) \\ \dot{k} = u \\ \dot{c} = 1 \\ \dot{p} = 0 \\ \dot{x_p} = 0}$};
		\node[state] at (10,0) (q_4) {\Large$\substack{\dot{x} = f(x,k) \\ \dot{k} = u \\ \dot{c} = 1 \\ \dot{p} = 0 \\ \dot{x_p} = 0}$};
		
		\path[->] (q_1)	edge [loop right] node[right] {\Large$\substack{(x \approx_{\epsilon} x_p \wedge c = p)? \\ c:=0 }$}();
		\path[->] (q_1)	edge [bend right=35] node[above=10] {\Large$\substack{(x \not\approx_{\epsilon} x_p \wedge c = p)? \\ c:=0 }$}(q_2);
		\path[->] (q_2) edge node[above] {\Large$\substack{(0<c \leq T_i)?}$} node[below] {\Large$\substack{c:=0 \\ x_p := x}$} (q_3);
		\path[->] (q_3) edge node[right] {\Large$\substack{(x \approx_{\epsilon} x_p \wedge 0 < c \le \delta)?}$} node[left] {\Large$\substack{p := c \\ c := 0 \\ x_p := x }$} (q_1);
		\path[->] (q_3) edge node[above] {\Large$\substack{(x \approx_{\epsilon} x_p \wedge c > \delta)?}$} node[below] {\Large$\substack{p:=c \\ c := 0 \\ x_p := x}$} (q_4);
		\path[->] (q_4)	edge [loop right] node[above=35, left=-2] {\Large$\substack{(x \approx_{\epsilon} x_p \wedge c = p)? \\ x_p := x \\ c:=0}$}();
		\path[->] (q_4) edge[bend left=65] node[above] {\Large$\substack{(x \not\approx_{\epsilon} x_p \wedge c = p)?}$} (q_2);
		
		\node at (3.5,3.7)  { $q_{STD}$};
		\node at (-1.8,-1.3)  { $q_{INIT}$};
		\node at (3.5,-1.3)  { $q_{LRN}$};
		\node at (11.5,-1.3)  { $q_{OSC}$};
		
		\end{tikzpicture} 
	\end{center}
	\caption{An oscillation property automaton.}
	\label{fig:osc_ha}
\end{figure}
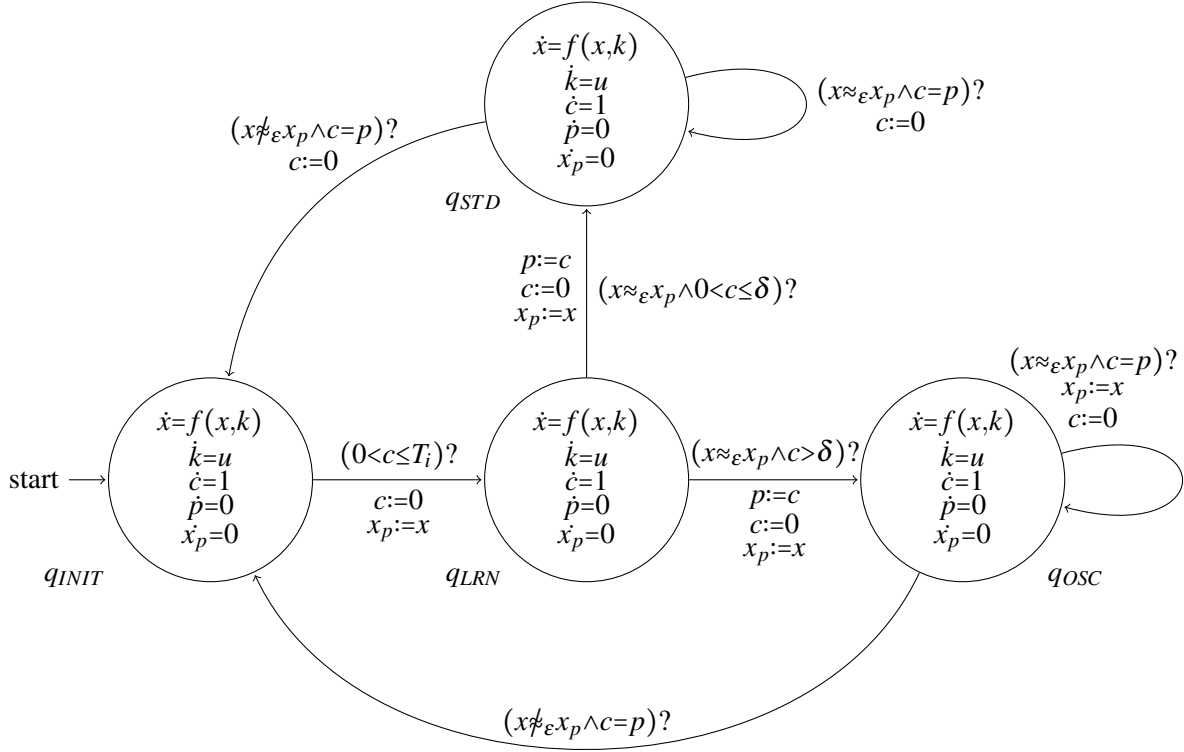

Suppose that we are interested in checking whether a given dynamical system exhibits the following behavioral pattern: the system from a given initial state evolves to a limit cycle and then under a some admissible parameter perturbation it evolves only from one limit cycle to another one. In other words, this parameter perturbation does not make the system undergo a structural behavior change (or a bifurcation). Another question is to know under which parameter changes the system moves from an oscillating behavior to a steady state.

The hybrid automaton depicted in Figure~\ref{fig:osc_ha} can be used to specify the above-described oscillating behaviors. In this property automaton, the parameters $k \in \real^m$ form part of the continuous state of the automaton. In the remainder of this paper, we assume that the evolution of $x$ can be described by:
\begin{equation*}
\begin{split}
\dot{x}  = &~f(x, k)\\
\dot{k}  = &~u \\
\end{split}
\end{equation*}
where $u$ is the input. This can be thought of as an abstract view of the dynamics of $x$, which can be described using complex concrete models, such as a hybrid automaton. Additionally, as we will show later, for test generation purposes, the continuous dynamics in the property automaton is abstracted away, which results in a discrete abstraction. This abstraction retains only important information about the expected temporal behavioral patterns of the variables under study. 


In this example, we restrict the derivatives of the parameters to be constant and they can take values in some set $U$. Therefore, this allows capturing piecewise linear evolution of the parameters. It is also worth noting that one can use other classes of functions to describe the parameter change. 

In addition, to describe the desired temporal behavioral pattern, we augment the continuous state of the property automaton with three special variables: $c$, $p$ and $x_p$, where $c$ is a clock used to measure time lapses, $p$ is used to store the oscillation period, and $x_p$ is used to memorize a point to which the system should return after a period.

The discrete structure of the property automaton consists of four locations $q_{INIT}$, $q_{LRN}$, $q_{OSC}$ and $q_{STD}$. The location $q_{INIT}$ corresponds to transient behaviors (between different qualitative behaviors) that can have a maximal duration $T_i$. After this amount $T_i$ of transient time, the automaton jumps to $q_{LRN}$ and while doing so, as specified by the associated reset map, it stores the current value of $x$ in the variable $x_p$ which is used as an expected periodic point.

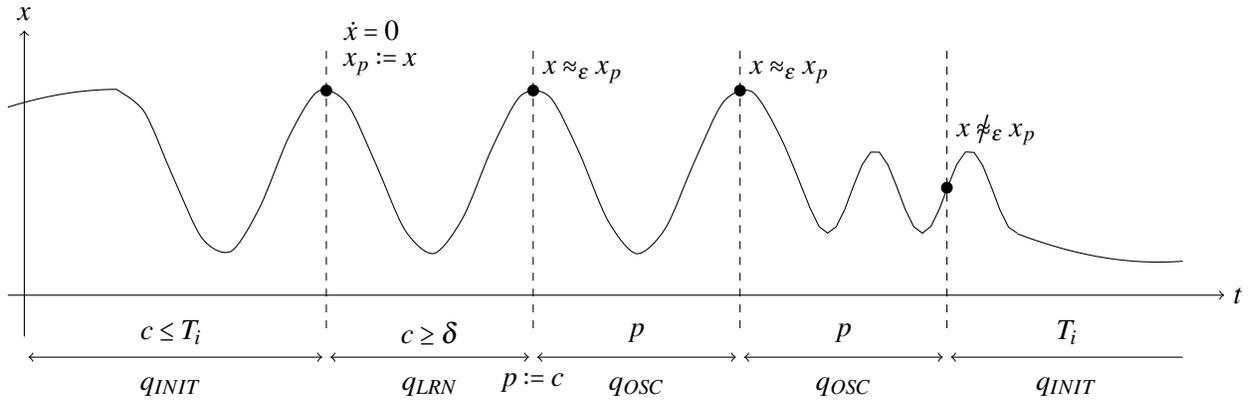
\begin{figure}
	\begin{center}
	\begin{tikzpicture}[scale=1.1]
	
		\draw[->] (-0.2,0) -- (14.5,0) node[right] {$t$}; 
		\draw[->] (0,-0.5) -- (0,3.2) node[above] {$x$}; 
		
		\draw[smooth, black, domain=1.8:3.1] plot (\x-2,{sin(1/2*\x r) +1.49});
		\draw[smooth, black, domain=3.1:11.5] plot (\x-2,{sin(2.5*\x r) +1.5});
		\draw[black, domain=11.5:14] plot (\x-2.0,{-1/2*sin(5.5*\x r) +1.25});
		\draw[smooth, black, domain=14:16] plot (\x-2,{-sin(1/2*\x r) +1.4});
		
		\draw[black, dashed, domain=-0.4:3] plot (3.65,{\x});
		\draw[black, dashed, domain=-0.4:3] plot (6.15,{\x});
		\draw[black, dashed, domain=-0.4:3] plot (8.65,{\x});
		\draw[black, dashed, domain=-0.4:3] plot (11.15,{\x});
		
		\draw[<->] (0.05,-0.75) -- (3.6,-0.75);
		\draw[<->] (3.7,-0.75) -- (6.1,-0.75);
		\draw[<->] (6.2,-0.75) -- (8.6,-0.75);
		\draw[<->] (8.7,-0.75) -- (11.1,-0.75);
		\draw[<-] (11.2,-0.75) -- (14,-0.75);
		
		\node at (1.775,-0.45)  { $c \leq T_i$};
		\node at (4.9,-0.45)  { $c \geq \delta$};
		\node at (6.14,-1.05)  { $p:=c$};
		\node at (7.4,-0.45)  { $p$};
		\node at (9.9,-0.45)  { $p$};
		\node at (12.6,-0.45)  { $T_i$};		
		
		\node at (1.775,-1.1)  { $q_{INIT}$};
		\node at (4.9,-1.1)  { $q_{LRN}$};
		\node at (7.4,-1.1)  { $q_{OSC}$};
		\node at (9.9,-1.1)  { $q_{OSC}$};
		\node at (12.6,-1.1)  { $q_{INIT}$};
		
		\draw[fill=black] (3.65,2.475) circle (0.065);
		\draw[fill=black] (6.15,2.475) circle (0.065);
		\draw[fill=black] (8.65,2.475) circle (0.065);
		\draw[fill=black] (11.15,1.3) circle (0.065);

		\node at (4.2,3.2)  { $\dot{x} = 0$};
		\node at (4.3,2.8)  { $x_p := x$};
		\node at (6.75,2.7)  { $x \approx_\epsilon x_p$};
		\node at (9.25,2.7)  { $x \approx_\epsilon x_p$};
		\node at (11.75,2)  { $x \not\approx_\epsilon x_p$};

	\end{tikzpicture}
	\end{center}
	\caption{Detection of an oscillation using a periodic point $x_p$. }
	\label{fig:osc_trace}
\end{figure}

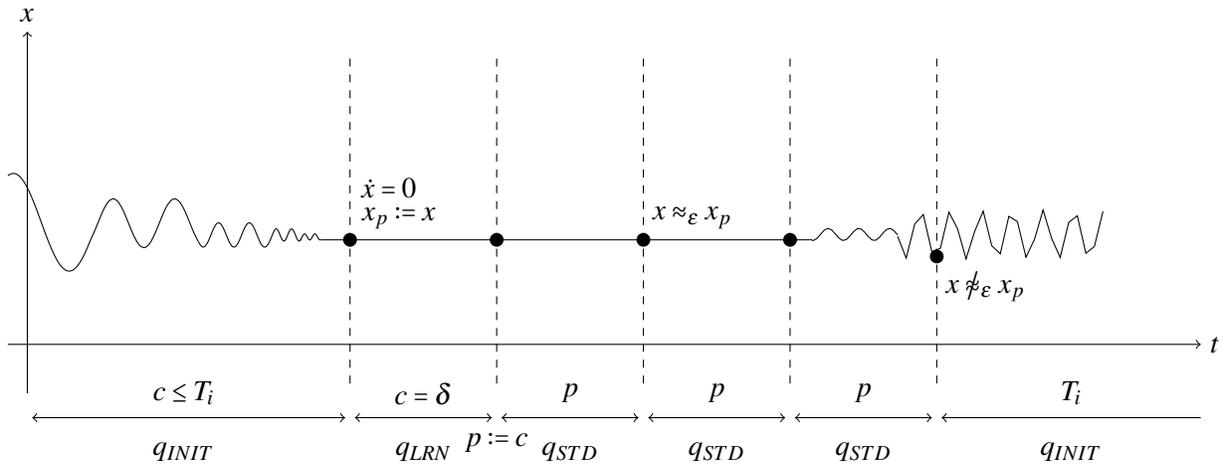
\begin{figure}
	\begin{center}
	\begin{tikzpicture}[scale=1.3]
	
		\draw[->] (-0.2,0) -- (12,0) node[right] {$t$}; 
		\draw[->] (0,-0.5) -- (0,3.2) node[above] {$x$}; 
		
		\draw[smooth, black, domain=0.8:1.65] plot (\x-1,{-1/2*sin(5.5*\x r) +1.25});
		\draw[smooth, black, domain=1.5:2.6] plot (\x-0.85,{-1/4*sin(10*\x r) +1.24});
		\draw[smooth, black, domain=2.7:3.45] plot (\x-0.95,{1/8*sin(20*\x r) +1.12});
		\draw[smooth, black, domain=3.45:3.75] plot (\x-0.95,{1/16*sin(40*\x r) +1.12});
		\draw[smooth, black, domain=3.75:3.96] plot (\x-0.96,{1/32*sin(60*\x r) +1.1});
		\draw[black, domain=4:9] plot (\x-1,{1.07});
		\draw[black, domain=9:9.9] plot (\x-1,{1/16*sin(20*\x r) +1.125});
		\draw[black, domain=9.9:12] plot (\x-1,{1/4*sin(20*\x r) +1.125});

		\draw[black, dashed, domain=-0.4:3] plot (3.3,{\x});
		\draw[black, dashed, domain=-0.4:3] plot (4.8,{\x});
		\draw[black, dashed, domain=-0.4:3] plot (6.3,{\x});
		\draw[black, dashed, domain=-0.4:3] plot (7.8,{\x});
		\draw[black, dashed, domain=-0.4:3] plot (9.3,{\x});
		
		\draw[<->] (0.05,-0.75) -- (3.25,-0.75);
		\draw[<->] (3.35,-0.75) -- (4.75,-0.75);
		\draw[<->] (4.85,-0.75) -- (6.25,-0.75);
		\draw[<->] (6.35,-0.75) -- (7.75,-0.75);
		\draw[<->] (7.85,-0.75) -- (9.25,-0.75);
		\draw[<-] (9.35,-0.75) -- (12,-0.75);
		
		\node at (1.6,-0.5)  { $c \leq T_i$};
		\node at (4.05,-0.5)  { $c = \delta$};
		\node at (4.8,-1)  { $p:=c$};
		\node at (5.55,-0.5)  { $p$};
		\node at (7.05,-0.5)  { $p$};		
		\node at (8.55,-0.5)  { $p$};
		\node at (10.675,-0.5)  { $T_i$};
		
		\node at (1.6,-1.1)  { $q_{INIT}$};
		\node at (4.05,-1.1)  { $q_{LRN}$};
		\node at (5.55,-1.1)  { $q_{STD}$};
		\node at (7.05,-1.1)  { $q_{STD}$};
		\node at (8.55,-1.1)  { $q_{STD}$};
		\node at (10.675,-1.1)  { $q_{INIT}$};
		
		\draw[fill=black] (3.3,1.07) circle (0.065);
		\draw[fill=black] (4.8,1.07) circle (0.065);
		\draw[fill=black] (6.3,1.07) circle (0.065);
		\draw[fill=black] (7.8,1.07) circle (0.065);
		\draw[fill=black] (9.3,0.9) circle (0.065);%
		
		\node at (3.7,1.6)  { $\dot{x} = 0$};
		\node at (3.8,1.3)  { $x_p := x$};
		\node at (6.8,1.3)  { $x \approx_\epsilon x_p$};
		\node at (9.8,0.6)  { $x \not\approx_\epsilon x_p$};

	\end{tikzpicture}
	\end{center}
	\caption{Detection of a steady state.}
	\label{fig:steady_trace}
\end{figure}

The role of the location $q_{LRN}$ is to ``learn'' the period of a limit cycle that the system is expected to enter. At location $q_{LRN}$, if after a strictly positive time $\delta$ the system returns to the point $x_p$, then the automaton resets the clock $c$ after storing its value in the variable $p$. We use a strictly positive amount of time lapse here to exclude Zeno behaviors. Therefore, if the system has entered a limit cycle, the value of the variable $p$ is exactly the period of that limit cycle (see Figure~\ref{fig:osc_trace}). In case the system reaches $x_p$ after exactly $\delta$ time, the automaton switches to the location $q_{STD}$ which is used to model a steady state (see Figure~\ref{fig:steady_trace}). Note that the test generation algorithm interacts with the system under test in discrete time, and the value of $\delta$ represents the smallest clock period that the test generation algorithm can handle.

After the learning phase at the location $q_{LRN}$ the variable $p$ contains the value of the expected period. When the automaton switches from the location $q_{LRN}$ to the location $q_{OSC}$, the variable $x_p$ is updated with the current value of $x$. At the location $q_{OSC}$, the automaton checks after every $p$ time whether $x$ returns to the periodic point $x_p$. There are two cases:
\begin{itemize}
\item If $x$ is in the $\epsilon$-neighborhood of the periodic point $x_p$, the system is considered oscillating and the clock is reset and the self-loop transition is traversed in order to check the next oscillation cycle. 
\item Otherwise, the automaton jumps back to the initial location $q_{INIT}$. This models the scenario where the system leaves the current limit cycle and may then evolve to another limit cycle.
\end{itemize}

To allow measurement imprecision, in the guard conditions $x$ is not required to return exactly to $x_p$ but to some $\epsilon$-neighborhood of $x_p$. This is denoted by $x \approx_{\epsilon} x_p$. 

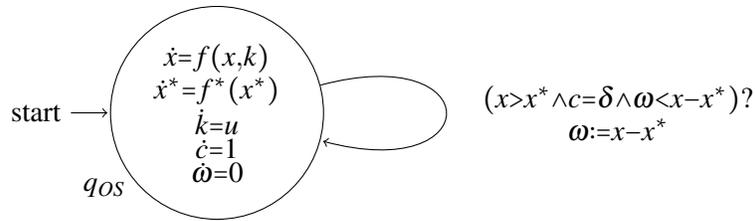
\begin{figure}
	\centering
		\begin{tikzpicture}[shorten >=1pt,node distance=2.5cm, auto]
			
		\node[state,initial] at (0,0) (q_1) {\Large $\substack{\dot{x} = f(x,k) \\ \dot{x}^* = f^*(x^*) \\ \dot{k} = u \\ \dot{c} = 1 \\ \dot{\omega} = 0 }$};		
		\path[->] (q_1)	edge [loop right] node[right=10] {\Large $\substack{(x > x^* \wedge c = \delta \wedge \omega < x-x^*)? \\ \omega:= x-x^*}$}();
		
		\node at (-1.5,-1)  {$q_{OS}$};
		
		\end{tikzpicture} 
		\caption{Overshoot detection.}
		\label{fig:shock_ha}
\end{figure}

Another property that can be easily expressed using hybrid automata is the maximal \emph{overshoot},
that is the maximum peak value of $x$ measured from a desired response of the system. The automaton has only one location $q_{OS}$ and is equipped with an auxiliary variable $\omega$ which stores the maximum distance from the desired response $x^*$ (see Figure~\ref{fig:shock_ha}).

\section{Guided Exploration to Falsify a Property}\label{sec:guided_exploration}
Given a property automaton, our problem now is to explore the parameter space to detect behaviors that do not satisfy the property expressed by this automaton. To do so, we make use of the test generation algorithm gRRT \cite{DangNahhalFMSD2009}. This algorithm is based on the star discrepancy coverage notion and allows achieving good coverage of the reachable state space. When the objective is not to cover the whole reachable space but to quickly detect some specific behavioral patterns, we can use on top of the gRRT algorithm a property-based guiding tool. The goal of this tool is to specify some critical regions to visit and then the algorithm gRRT can be used to cover those regions. Before continuing, let us briefly recall the algorithm gRRT.

Given a hybrid automaton $\A$, the algorithm gRRT generates a {\em test case} represented by a tree where each node is associated with a state of $\A$ and each arc is associated with a control input action, which is either a continuous input value or a discrete control action (that is, the action of traversing a transition). Note that in the context of this work, both continuous inputs (described by $u$ in the definition of hybrid automata) and transitions are controllable by the tester. To execute such a test case, the tester applies a control input sequence to the system, measures the variables of interest and decides whether the system under test satisfy the property. 
The algorithm thus can be thought as a procedure to find input signals that correspond to the beahviors we want to observe.
The main steps of the coverage-guided test generation algorithm gRRT \cite{DangNahhalFMSD2009} are the following: 
\begin{itemize}
\item Step 1: a goal state $(q_{goal}, \xgoal)$ is sampled from the state space; 
\item Step 2: a neighbor state $(q_{near}, \xnear)$ of the goal state is determined;
\item Step 3: from the neighbor state, an appropriate continuous input $u$ is applied for a time step $h$, or a transition is taken, in order to steer the system towards the goal state.
\end{itemize}
Step 2 can be done using a notion of distance between two hybrid states that capture the effects of discrete transitions. The choice of continuous input $u$ in Step 3 can be done by a random selection from a discretization of the input set $U$. Indeed, more sophisticated methods based on trajectory sensitivity to input variation can be used but they cost more computation effort. It is important to note that a good selection of goal states is key to a good coverage result, because the success of randomized algorithms depends on finding good starting states. For a more thorough description of the algorithm gRTT and its properties, the reader is referred to \cite{DangNahhalFMSD2009}. For the example in Section~\ref{sec:application} a uniform randomized selection of $u$ is used, which already allows an efficient exploration.

To bias the goal state sampling while taking into account the property to falsify, we first construct a discrete abstraction of the property automaton $\A$ that reflects the expected behavioral patterns. This abstraction is then used to biased the goal state samplings, so that it favors the exploration of the behavioral patterns of interest. As an example, to falsify the oscillation property presented in the previous section, that is to show that after a given initial transient time there exists a parameter change that leads the system out of an expected limit cycle, the trajectories that lead to the transition from $q_{OSC}$ to $q_{INIT}$ are favored. In other words, the exploration is biased in a way to increase the probability of sampling the goal states in the guard set of this transition.

To define a discrete abstraction, we need some additional definitions. A $n$-dimensional predicate is defined as $\LinPred(x) :=  g(x) \sim 0$ where $g: \real^{n} \to \real$ is a function of $n$ variables, and $\sim \in \{ \ge, > \}$. Let $\SLinPred$ be a function that specifies for each location $q \in Q$ a vector of $m_q$ predicates, that is $\SLinPred(q) = (\LinPred_1, \ldots, \LinPred_{m_q})$. 

We define for each location $q$ an abstraction function $\absFct_q: \X \to \bool^{m_q}$ such that 
$$\absFct_q(x) = (\LinPred_1(x), \ldots, \LinPred_{m_q}(x)).$$ 
We say that the Boolean abstraction vector of $x$ with respect to $\absFct_q$ is the Boolean vector $(\LinPred_1(x), \ldots, \LinPred_{m_q}(x))$. The abstraction function $\absFct_q$ associated with a location $q \in Q$ partitions the set of continuous states at location $q$ into at most $2^{m_q}$ subsets of continuous states such that all the continuous states in each subset have the same Boolean abstraction vector with respect to the abstraction function $\absFct_q$. 

In the other direction, for each location $q$ we define the concretization function $\concFct_q: \bool^{m_q} \to 2^{\X}$ such that for a given Boolean vector $b \in \bool^{m_q}$, $\concFct_q(b) = \{ x \in \X ~|~ \absFct_q(x) = b \}$.

The discrete abstraction of $\A$ with respect to $\SLinPred$ is a transition system $\dA = \{ \abStSpace, \absTrans, \absst_0 \}$. 
\begin{itemize}
\item Each location $q$ of the hybrid automaton corresponds to a set $S_q$ of abstract states, each of which corresponds to a pair $(q, b)$ where $b \in \bool^{m_q}$ is a value of the Boolean abstraction vector. For convenience, we call them $q$-abstract states. Two $q$-abstract states $\absst = (q, b)$ and $\absst'=(q,b')$ are adjacent if their corresponding sets of concrete states, that is $\concFct_q(b)$ and $\concFct_q(b')$, have non-empty intersection and they intersect only their boundaries. The whole abstract state space $\abStSpace$ is the union 
$$\abStSpace = \bigcup_{q \in Q} S_q.$$ 
\item The transition relation $\absTrans \subset \abStSpace \times \abStSpace$ between the abstract states is defined as the union of the following two relations $\absTrans_d$ and $\absTrans_c$. Let $\absst = (q, b)$ and $\absst'=(q',b')$ be two abstract states; the transition relation between them is defined as follows:
\begin{itemize}
\item $\absst \absTrans_c \absst'$ if $q=q'$ and $\absst_1$ and $\absst_2$ are adjacent.
\item $\absst \absTrans_d \absst'$ if $q \neq q'$ and $\concFct_q(\absst) \cap \Guard_{qq'} \neq \emptyset$ and $\mathcal{R}(\concFct_q(\absst) \cap \Guard_{qq'}) \se \mathcal{I}_{q'}$.
\end{itemize}

The relation $\absTrans_d$ represents the transitions in the abstract state space due to discrete switches in the original hybrid automaton $\A$, the relation $\absTrans_c$ represents the continuous evolution in $\A$.

\item The initial abstract state $\absst_0=(q_0, b_0)$ where $b_0=\absFct_{q_0}(x_0)$.
\end{itemize}

The abstraction $\dA$ can be thought of as an over-approximation of $\A$ since it is easy to see that any execution of $\A$ corresponds to an execution of $\dA$. Moreover, it can be refined based on the exploration results in order to distinguish different qualitative behaviors that are important with respect to the property to validate.

In order for such a discrete abstraction to reflect the behavioral patterns we want to explore, we should choose for each location a set of predicates that can capture the discrete transitions of $\A$ and separate critical regions from the rest; therefore the set should include the predicates defining the guard and invariant conditions. This will be illustrated by the example in Section~\ref{sec:application}.


To biased the search, we use the Metropolis-Hastings method to perform a random walk \cite{Motwani:1995:RA:211390} on $\dA$ starting at the abstract state $s_0$. We first specify a target probability distribution over the abstract states
$$\pi = \{ \pi_s ~|~ s \in \abStSpace \}.$$  
We then construct the following transition matrix $P(\dA)$. Between two abstract states $s$ and $s'$, we assign a probability to the transition from $s$ to $s'$:
$$\left \{ \begin{array}{ll}
         \displaystyle{p_{ss'} = \frac{1}{deg(s)} min \{ \frac{deg(s) \pi_{s'}}{deg(s') \pi_s}, 1 \}} & \text{if $s \absTrans s'$}  \\
         \displaystyle{p_{ss'} = 1 - \sum_{w \neq s} p_{sw}} & \text{if $s=s'$} \\
         \displaystyle{p_{ss'} = 0} & \text{otherwise} \\
\end{array} \right.$$
The above transition matrix $P(\dA)$ guarantees that the stationary distribution of the resulting random walk on the abstraction $\dA$ is the target distribution $\pi$ \cite{Nonaka20101889}.  Therefore the abstract states corresponding to the region we want to visit are assigned with high target probabilities.

The Metropolis-Hastings method was proved to have good hitting times, which allows quickly reaching a desired abstract state, indeed the hitting time from $s$ to $s'$ of this random walk is of $\mathcal{O}(rN_v^2)$ where $N_v$ is the number of abstract states and $r = \max \{ \displaystyle{\frac{\pi_s}{\pi_{s'}}} ~|~ s, s' \in \absTrans \}$. 


\section{Application}\label{sec:application}

\subsection{Laub-Loomis Model}

In this section we apply on the Laub-Loomis model~\cite{laub1998molecular} the techniques previously exposed. The model consists of a parametrized ODE system extracted from a molecular network that describes the aggregation stage of Dictyostelium. Our main intent is to show that for some parameter variation with bounded derivatives, the spontaneous oscillations of the system do not occur any more. Roughly speaking, we want to falsify the oscillation robustness of the system.

To this end, we derive a discrete abstraction from the property automaton in Figure~\ref{fig:osc_ha} and guide the simulation of the ODE system towards the areas in the state space of the property automaton where the oscillation disappears. The derivatives $u$ of the parameter variables are the inputs that we use to guide the exploration.

A revisited model that slightly differs from the original one presented by Laub and
Loomis~\cite{laub1998molecular} is the following~\cite{4434348}:

\begin{equation*}
	\dot{x} = f(x,k) = 
	\begin{bmatrix}
		k_1 x_7 - k_2 x_1 x_2 \\
		k_3 x_5 - k_4 x_2 \\
		k_5 x_7 - k_6 x_2 x_3 \\
		k_7 - k_8 x_3 x_4 \\
		k_9 x_1 - k_{10} x_4 x_5 \\
		k_{11} x_1 - k_{12} x_6 \\
		k_{13} x_6 - k_{14} x_7 \\
	\end{bmatrix}
	\qquad
	\begin{tabular}{| l l | l l |}
		\hline
		Par. & Val. & Par. & Val. \\
		\hline
		$k_1$	& $2.0$ $min^{-1}$	& $k_8$	& $1.3$ $min^{-1}$	\\
		$k_2$	& $0.9$ $min^{-1}$	& $k_9$	& $0.3$ $min^{-1}$	\\
		$k_3$	& $2.5$ $min^{-1}$	& $k_{10}$	& $0.8$ $min^{-1} \mu M^{-1}$	\\
		$k_4$	& $1.5$ $min^{-1}$	& $k_{11}$	& $0.7$ $min^{-1}$	\\
		$k_5$	& $0.6$ $min^{-1}$	& $k_{12}$	& $4.9$ $min^{-1}$	\\
		$k_6$	& $0.8$ $min^{-1} \mu M^{-1}$	& $k_{13}$	& $23.0$ $min^{-1}$	\\
		$k_7$	& $1.0$ $min^{-1} \mu M^{-1}$	& $k_{14}$	& $4.5$ $min^{-1} \mu M^{-1}$	\\[0.05cm]
		\hline
		\multicolumn{4}{c}{} \\[-0.3cm]
		\multicolumn{4}{c}{Table 1: Oscillations parameter values.} \\
	\end{tabular}
\end{equation*}

The variables $x$ correspond to seven protein concentrations:
$x_1 = [ACA]$, $x_2 = [PKA]$, $x_3 = [ERK2]$, $x_4 = [REGA]$, $x_5 = [Internal\ cAMP]$, $x_6 = [External\ cAMP]$
and $x_7 = [CAR1]$. The coefficient vector $k = [k_1, \dots, k_{14}]$ contains the system parameters. Table 1 shows the parameter values for which spontaneous oscillations occur~\cite{laub1998molecular}.

\subsection{Constructing a Discrete Abstraction} 

In the property automaton in Figure~\ref{fig:osc_ha}, the transition from $q_{OSC}$ to the location $q_{INIT}$ is critical since it takes the system from an oscillation phase to a non-oscillation phase. We thus want to control the system's behavior in order to satisfy the condition $(x \not \approx_\epsilon x_p)?$. 

\begin{figure}[htbp]
	\centering
	
	\begin{subfigure}[b]{0.45\textwidth}
		\centering
		\begin{tikzpicture}[shorten >=1pt,node distance=2cm, auto]
			
		\node[state] at (2,0) (q_4) {\Large $\substack{\dot{x} = f(x,k) \\ \dot{z} = f(x,k) \\ \dot{k} = u \\ \dot{c} = 1 \\ \dot{p} = \dot{x_p} = 0}$};
		\node at (0,-3) (q_2){};
		
		\path[->] (q_4)	edge [loop right] node[right] {\Large $\substack{(|z| \leq \epsilon \wedge c = p)? \\ z := 0 \\ c:=0}$}();
		\path[->] (q_4) edge[bend left=75] node[above=7pt,left=5pt] {\Large $\substack{(|z| > \epsilon \wedge c= p)?}$} (q_2);

		\end{tikzpicture} 
		\caption{$q_{OSC}$ modified.}
		\label{fig:mod_q_osc}
	\end{subfigure}
	\begin{subfigure}[b]{0.45\textwidth}
		\centering
		\centering
		\begin{tikzpicture}[shorten >=1pt,node distance=2cm, auto]
			
		\node[state] at (2,0) (q_1) {\Large $\substack{\dot{x} = f(x,k) \\ \dot{z} = f(x,k) \\ \dot{k} = u \\ \dot{c} = 1 \\ \dot{p} = \dot{x_p} = 0}$};
		\node at (0,-3) (q_2){};
		
		\path[->] (q_1)	edge [loop right] node[right] {\Large $\substack{(|z| \leq \epsilon \wedge c = p)? \\ c:=0 }$}();
		\path[->] (q_1)	edge [bend right=35] node[left=5] {\Large $\substack{(|z| > \epsilon \wedge c = p)? \\ c:=0 }$}(q_2);
		
		\end{tikzpicture} 
		\caption{$q_{STD}$ modified.}
		\label{fig:mod_q_std}
	\end{subfigure}
	\caption{Modified property automaton.}
\end{figure}
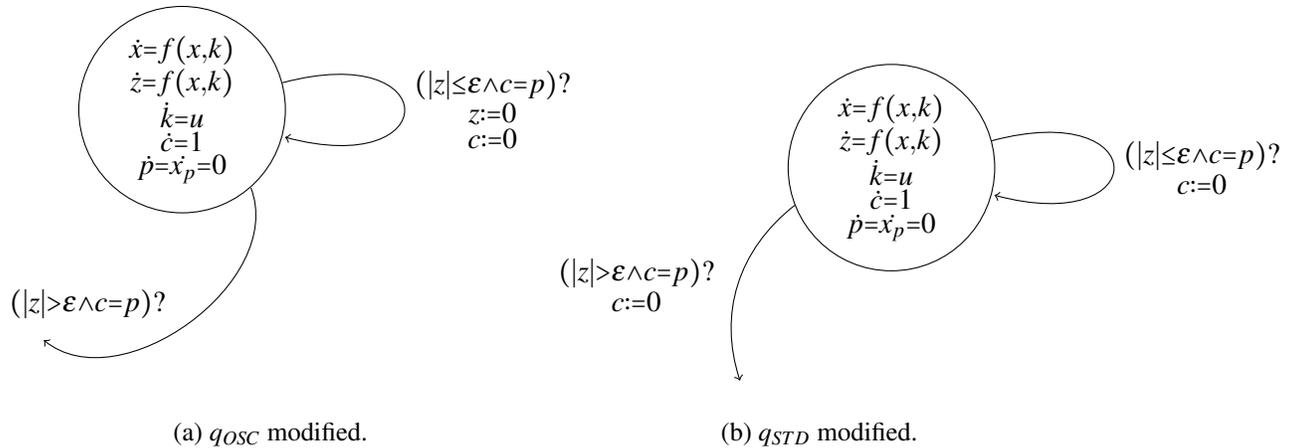

In addition, we modify the property automaton so that it results in an abstraction with predicates involving only one state variable, which is more suitable for the algorithm gRRT. Indeed the star discrepancy is defined for states inside some rectangular sets; for more general sets, box approximations are required. To do so, we modify the condition $(x \approx_\epsilon x_p)?$ by introducing a new variable $z = x - x_p$, the  
derivative of which is $\dot{z} = \dot{x} - \dot{x_p} = \dot{x}$ (recall that by definition $\dot{x_p} = 0$). The guard on the self-loop transition over $q_{OSC}$ becomes
$(x \approx_\epsilon x_p)? \equiv (| x- (x - z) | < \epsilon)? \equiv (| z | \leq \epsilon)?$, while the reset is 
rewritten as $(x_p := x) \equiv (x-z := x) \equiv (z := 0)$. Similarly the guard condition of the transition from $q_{OSC}$ to $q_{INIT}$ becomes $(|z| > \epsilon)?$ and $z := 0$, respectively (see Figure~\ref{fig:mod_q_osc}). The guard conditions and resets concerning the clock $c$ remain unchanged. The same reasoning can be easily applied to the location $q_{STD}$ (see Figure~\ref{fig:mod_q_std}).

We now proceed with the definition of the function $\SLinPred$ which is the basis of the abstraction. Let 
$\SLinPred : Q \rightarrow 2^{\Pi}$ be defined as follows:
\begin{equation*}
	\lambda(q) = \begin{cases}
		( z \geq \epsilon, \epsilon > z, z > -\epsilon, -\epsilon \geq z ) & \text{if $q = q_{OSC}$,} \\
		(\top)		&	\text{otherwise.}
	\end{cases}
\end{equation*}
Note that abstraction function partitions the space of $z$ in $q_{OSC}$ into the sets $(+\infty;+\epsilon]$, $(+\epsilon,-\epsilon)$ and $[-\epsilon,-\infty)$ with the respective 
Boolean abstraction vectors $(1,0,0,0)$, $(0,1,1,0)$ and $(0,0,$ $0,1)$. From $\SLinPred$ we obtain the transition system $\dA = \{ \abStSpace, \absTrans, \absst_0 \}$,
with abstract states
\begin{eqnarray*}
	\abStSpace = 	& \{ s_i = (q_i,\top) \mid q_i \in Q . q_i  \neq q_{OSC} \} \cup \\
				& \{ s'_{OSC} = (q_{OSC},(1,0,0,0)) ; s''_{OSC} = (q_{OSC},(0,1,1,0)) ; s'''_{OSC} = (q_{OSC},(0,0,0,1)) \},
\end{eqnarray*}

transition relation 
\begin{eqnarray*}
		\absTrans = 	& \absTrans_C \cup \absTrans_D = \\
					& \{ (s'_{OSC},s''_{OSC}) ; (s''_{OSC},s'_{OSC}); (s''_{OSC},s'''_{OSC}); (s'''_{OSC},s''_{OSC})  \} \cup \\
					& \{ ((q,b) , (q',b') ) \mid  (q,q') \in E . q,q' \neq q_{OSC}\} \cup \\ 
					& \{ (s'_{OSC},s_{INIT}) ; (s''_{OSC},s''_{OSC}) ; (s'''_{OSC},s_{INIT}) \}, \\
\end{eqnarray*}
and the initial abstract state $s_0 = s_{INIT} = (q_{INIT}, \top)$.

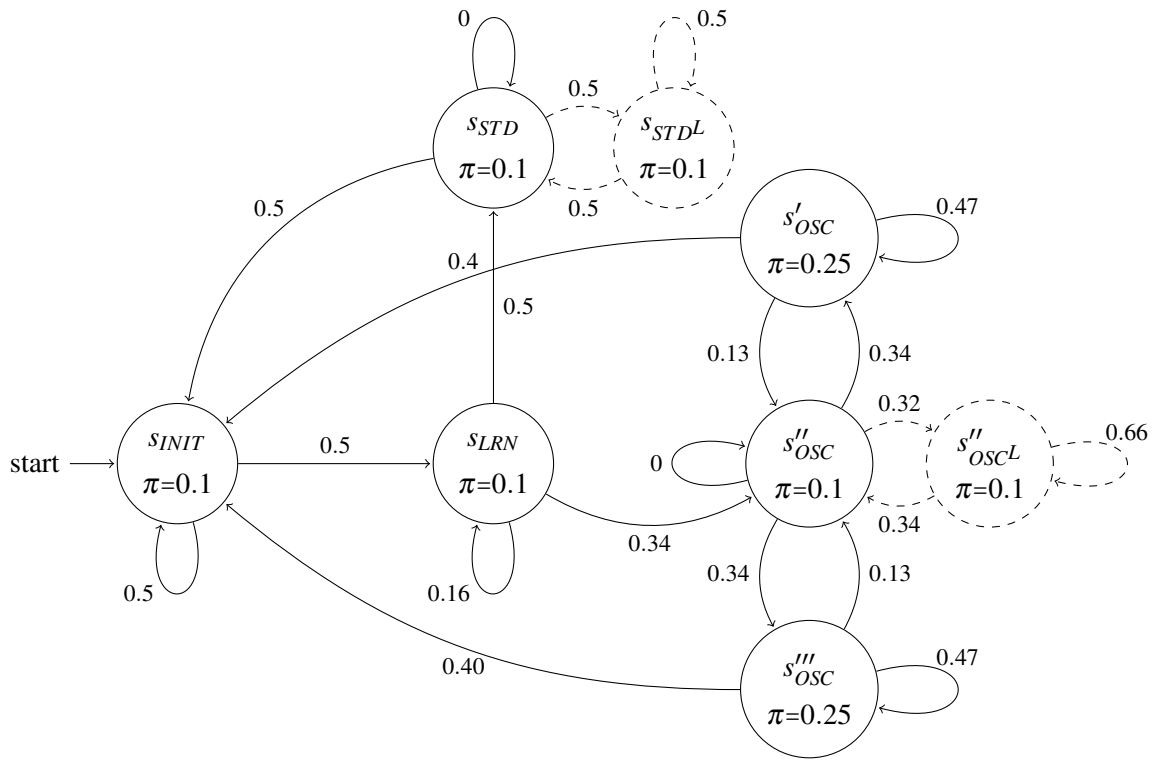
\begin{figure}[htbp]
	\begin{center}
		\begin{tikzpicture}[shorten >=1pt,node distance=2cm, auto, scale=1.2]
	
		\node[state] at (3.5,3.5) (q_1) {\Large $\substack{s_{STD} \\ \\ \pi = 0.1 }$};
		\node[state, dashed] at (5.5,3.5) (ql_1) {\Large $\substack{s_{STD^L} \\ \\ \pi = 0.1 }$};
		\node[state,initial] at (0,0) (q_2) {\Large $\substack{s_{INIT} \\ \\ \pi = 0.1 }$};
		\node[state] at (3.5,0) (q_3) {\Large $\substack{s_{LRN} \\ \\ \pi = 0.1 }$};
		\node[state] at (7,2.5) (qp_4) {\Large $\substack{s'_{OSC} \\ \\ \pi = 0.25 }$};
		\node[state] at (7,0) (qs_4) {\Large $\substack{s''_{OSC} \\ \\ \pi = 0.1 }$};
		\node[state, dashed] at (9,0) (qsl_4) {\Large $\substack{s''_{OSC^L} \\ \\ \pi = 0.1 }$};
		\node[state] at (7,-2.5) (qt_4) {\Large $\substack{s'''_{OSC} \\ \\ \pi = 0.25 }$};
		
		\path[->] (q_1)	edge [bend left, dashed] node[above] {\footnotesize $0.5$} (ql_1);
		\path[->] (q_1)	edge [bend right=35] node[above=5] {\footnotesize $0.5$}(q_2);
		\path[->] (q_1)	edge [loop above] node[left=5] {\footnotesize $0$}();
		\path[->] (ql_1)	edge [bend left, dashed] node[below] {\footnotesize $0.5$} (q_1);
		\path[->] (ql_1)	edge [loop above, dashed] node[right=5] {\footnotesize $0.5$} ();
		\path[->] (q_2) edge node[above] {\footnotesize $0.5$} (q_3);
		\path[->] (q_2) edge [loop below] node[left=5] {\footnotesize $0.5$} ();
		\path[->] (q_3) edge node[right] {\footnotesize $0.5$} (q_1);
		\path[->] (q_3) edge [bend right] node[below] {\footnotesize $0.34$} (qs_4);
		\path[->] (q_3) edge[loop below] node[left=5] {\footnotesize $0.16$} ();
		\path[->] (qp_4) edge[bend right=20] node[above] {\footnotesize $0.4$} (q_2);
		\path[->] (qp_4) edge[bend right] node[left] {\footnotesize $0.13$} (qs_4);
		\path[->] (qp_4) edge[loop right] node[above=5] {\footnotesize $0.47$} ();
		\path[->] (qs_4)	edge [bend left, dashed] node[above] {\footnotesize $0.32$}(qsl_4);
		\path[->] (qs_4)	edge [bend right] node[right] {\footnotesize $0.34$} (qp_4);
		\path[->] (qs_4)	edge [bend right] node[left] {\footnotesize $0.34$} (qt_4);
		\path[->] (qs_4)	edge [loop left=20] node[left] {\footnotesize $0$} ();
		\path[->] (qsl_4) edge [bend left, dashed] node[below] {\footnotesize $0.34$}(qs_4);
		\path[->] (qsl_4) edge [loop right, dashed] node[above=5] {\footnotesize $0.66$}();
		\path[->] (qt_4)	edge [bend right] node[right] {\footnotesize $0.13$} (qs_4);
		\path[->] (qt_4)	edge [bend left=20] node[below] {\footnotesize $0.40$} (q_2);
		\path[->] (qt_4)	edge [loop right] node[above=5] {\footnotesize $0.47$} ();
		
		\end{tikzpicture} 
	\end{center}
	\caption{Abstract transition system of the property automaton. Dashed
	states and transitions are introduced to eliminate self-loop transitions in the property automaton.}
	\label{fig:abs_osc}
\end{figure}

Before specifying the target probabilities over the abstract states, it is necessary to make another modification to the abstract transition system, in order to be able to distinguish the self-loop transitions originated from the abstraction process from those introduced by the transition probability definition. This modification consists in duplicating the locations that with self-loop transitions and replacing these self-loop transitions with the transitions  connecting the original location to its copy, and
vice versa. Hence, for this example we add two locations $s_{STD^L}$ and $s''_{OSC^L}$ to $\abStSpace$ and we replace in $\absTrans$ the transitions $(s_{STD},s_{STD})$ and $(s''_{OSC},s''_{OSC})$ with $(s_{STD},s_{STD^L})$, $(s_{STD^L},s_{STD})$, $(s''_{OSC},s''_{OSC^L})$ and $(s''_{OSC^L},s''_{OSC})$. Figure~\ref{fig:abs_osc} shows the resulting actraction without such self-loop transitions. 

We now define the target probabilities over the abstract states. Since we are interested in detecting that the system stops oscillating, it makes sense to attribute higher probabilities 
to those abstract states which bring the system from an oscillation phase to a non-oscillation one, i.e., the states $s'_{OSC}$ and $s'''_{OSC}$. Thus, defining the target probabilities as 
$\pi_{s_{INIT}} = \pi_{s_{LRN}} = \pi_{s_{STD}} = \pi_{s_{STD^L}} = \pi_{s''_{OSC}} = \pi_{s''_{OSC^L}} = 0.1$ and $ \pi_{s'_{OSC}} = \pi_{s'''_{OSC}} = 0.25$,
we obtain the following probability transition matrix:
\begin{center}
	\begin{tabular}{c | c | c | c | c | c | c | c | c | c c}
					&	$s_{INIT}$	&	$s_{LRN}$	&	$s_{STD}$	&	$s_{STD^L}$	&	$s'_{OSC}$	&	$s''_{OSC}$	&	$s''_{OSC^L}$	&	$s'''_{OSC}$ 	& \\[0.5pt]
		\hline
		$s_{INIT}$	&		0.50		&		0.50		&		0		&		0		&		0		&		0		&		0		&		0		& \\
		$s_{LRN}$	&		0		&		0.16		&		0.50		&		0		&		0		&		0.34		&		0		&		0		& \\
		$s_{STD}$	&		0.50		&		0		&		0		&		0.50		&		0		&		0		&		0		&		0		& \\
		$s_{STD^L}$	&		0		&		0		&		0.50		&		0.50		&		0		&		0		&		0		&		0		& \\
		$s'_{OSC}$	&		0.40		&		0		&		0		&		0		&		0.47		&		0.13		&		0		&		0		& \\
		$s''_{OSC}$	&		0		&		0		&		0		&		0		&		0.34		&		0		&		0.32		&		0.34		& \\
		$s''_{OSC^L}$	&		0		&		0		&		0		&		0		&		0		&		0.34		&		0.66		&		0		& \\
		$s'''_{OSC}$	&		0.40		&		0		&		0		&		0		&		0		&		0.13		&		0		&		0.47		& \\
		\end{tabular}
\end{center}
that leads to the system shown in Figure~\ref{fig:abs_osc}.

\subsection{Experimental Results}\label{sec:ex_results}

	We have implemented the above described method and incorporated it in the HTG tool~\cite{dang_model_based}, which is
	our previous C++ implementation of the gRRT algorithm. In particular, we extended HTG
	with the following new functions: defining a discrete abstraction over the considered hybrid automaton,
	specifying the target probabilities for each abstract state and performing a random walk on the abstract transition system in order to identify the areas that need to be explored.
	
	In our experiments, we focused on the parameter $k_1$ and on its derivative modelled by the input variable $u_1$. Moreover,
	we monitor the two variables $x_1$ and $z_1$ since $k_1$ is involved in both of their dynamics.
	The values of the other parameters of the automaton in Figure~\ref{fig:osc_ha} are fixed as follows:
	$T_i = 7.3781$, $\delta = 0.05$ and $\epsilon = 0.2$. As an initial value of $k_1$ we choose its
	oscillating nominal value $2.0$ (see Table 1).

\begin{figure}
		\centering
		\includegraphics[scale=0.7]{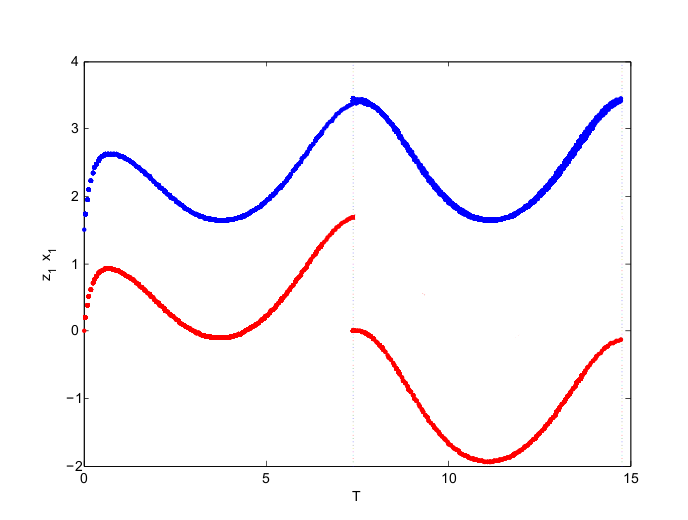}
	
		\caption{Evolutions of $x_1$ (blue) and $z_1$ (red) for $u_1 \in [-0.01,0.01]$}
		\label{fig:u001}
	\end{figure}

\begin{figure}
		\centering
		\includegraphics[scale=0.7]{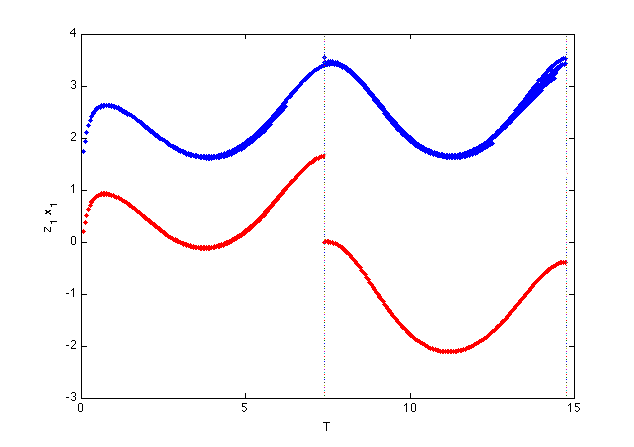}
	
		\caption{Evolutions of $x_1$ (blue) and $z_1$ (red) for $u_1 \in [-0.1,0.1]$}
		\label{fig:u01}
	\end{figure}

	We performed three experiments with different ranges of the input $u_1$. In the first case $u_1$ can be sampled within the interval $[-0.01, 0.01]$ (see Figure~\ref{fig:u001}),
	in the second within $[-0.1,0.1]$ (see Figure~\ref{fig:u01}), while in the third within [-1.0,1.0]. In all the experiments the state space
	of $k_1$ is $[1.8,2.2]$.
	In the first case, even if at the end of each period the value of $z_1$ is
	not exactly equal to zero, it is always included in the interval defined by $\epsilon = 0.2$ and thus, for all the 
	simulation runs, the system is considered oscillating. Differently, in the case where $k_1$ can evolve faster, the variable $z_1$ ends an oscillation phase at a value smaller than 
	$-\epsilon$. This means that already for values of
	$k_1 \in [1.8,2.2]$ and $\dot{k_1} = u_1 \in [-0.1,0.1]$ the system leaves the current limit cycle.
	We can interpret such a behavioral change under a small variation of the nominal parameter values as weak robustness of the Laub-Loomis model. Finally, for values of $u_1 \in [-1.0,1.0]$ we 
	found that, not very surprisingly, the variable $z_1$ drifts very far away from zero, 
	showing that the system has stopped oscillating.
	
	All the experiments were performed on a MacBook $3,1$ having $2$GB RAM. Each experiment 
	involved the computation of $30000$ points, with integration time step equal to $0.05$.
	In the first experiment, the tool required $6.23s$, in the second $5.94s$ and in the 
	third $6.53s$. To give an idea of the scalability of the technique,
	a simulation with $10000$ points requires $1.22s$, 
	with $25000$ points $4.93s$, while with $50000$ points $17.42s$.

\vspace{-0.5cm}
\section{Conclusion}\label{sec:conclusion}
\vspace{-0.5cm}
In this paper, we described a framework for falsifying oscillation properties and study the robustness of biological models. The experimental results are encouraging and we intend to pursue this work in two directions. One is a more efficient parameter sampling, which can be guided by local analysis using Floquet theory. The other direction concerns the application of this approach to analyze other types of bifurcation in biological systems.

\bibliographystyle{eptcs}
\bibliography{hybridbib}

\begin{thebibliography}{10}
\providecommand{\bibitemdeclare}[2]{}
\providecommand{\surnamestart}{}
\providecommand{\surnameend}{}
\providecommand{\urlprefix}{Available at }
\providecommand{\url}[1]{\texttt{#1}}
\providecommand{\href}[2]{\texttt{#2}}
\providecommand{\urlalt}[2]{\href{#1}{#2}}
\providecommand{\doi}[1]{doi:\urlalt{http://dx.doi.org/#1}{#1}}
\providecommand{\bibinfo}[2]{#2}

\bibitemdeclare{inproceedings}{Alur}
\bibitem{Alur}
\bibinfo{author}{R.~\surnamestart Alur\surnameend},
  \bibinfo{author}{C.~\surnamestart Courcoubetis\surnameend},
  \bibinfo{author}{T.~A. \surnamestart Henzinger\surnameend} \&
  \bibinfo{author}{P.~H. \surnamestart Ho\surnameend} (\bibinfo{year}{1992}):
  \emph{\bibinfo{title}{Hybrid {A}utomata: {A}n {A}lgorithmic {A}pproach to the
  {S}pecification and {V}erification of {H}ybrid {S}ystems}}.
\newblock In \bibinfo{editor}{R.~L. \surnamestart Grossman\surnameend},
  \bibinfo{editor}{A.~\surnamestart Nerode\surnameend}, \bibinfo{editor}{A.~P.
  \surnamestart Ravn\surnameend} \& \bibinfo{editor}{H.~\surnamestart
  Richel\surnameend}, editors: {\sl \bibinfo{booktitle}{Hybrid Systems}},
  \bibinfo{series}{LNCS}, \bibinfo{publisher}{Springer}, pp.
  \bibinfo{pages}{209--229}, \doi{10.1007/3-540-57318-6\_30}.

\bibitemdeclare{article}{alur94}
\bibitem{alur94}
\bibinfo{author}{R.~\surnamestart Alur\surnameend} \& \bibinfo{author}{D.~L.
  \surnamestart Dill\surnameend} (\bibinfo{year}{1994}):
  \emph{\bibinfo{title}{A Theory of Timed Automata}}.
\newblock {\sl \bibinfo{journal}{Theoret. Comput. Sci.}}
  \bibinfo{volume}{126}(\bibinfo{number}{2}), pp. \bibinfo{pages}{183--235},
  \doi{10.1016/0304-3975(94)90010-8}.

\bibitemdeclare{article}{Bartocci2009}
\bibitem{Bartocci2009}
\bibinfo{author}{E.~\surnamestart Bartocci\surnameend},
  \bibinfo{author}{F.~\surnamestart Corradini\surnameend},
  \bibinfo{author}{E.~\surnamestart Merelli\surnameend} \&
  \bibinfo{author}{L.~\surnamestart Tesei\surnameend} (\bibinfo{year}{2009}):
  \emph{\bibinfo{title}{Model Checking Biological Oscillators}}.
\newblock {\sl \bibinfo{journal}{Electronic Notes in Theoretical Computer
  Science}} \bibinfo{volume}{299}(\bibinfo{number}{1}), pp.
  \bibinfo{pages}{41--58}, \doi{10.1016/j.entcs.2009.02.004}.

\bibitemdeclare{book}{dang_model_based}
\bibitem{dang_model_based}
\bibinfo{author}{T.~\surnamestart Dang\surnameend} (\bibinfo{year}{2010}):
  \emph{\bibinfo{title}{Model-based testing of hybrid systems}}.
\newblock \bibinfo{series}{Model-Based Testing for Embedded Systems},
  \bibinfo{publisher}{CRC Press}, \doi{10.1201/b11321-15}.

\bibitemdeclare{article}{DangNahhalFMSD2009}
\bibitem{DangNahhalFMSD2009}
\bibinfo{author}{T.~\surnamestart Dang\surnameend} \&
  \bibinfo{author}{T.~\surnamestart Nahhal\surnameend} (\bibinfo{year}{2009}):
  \emph{\bibinfo{title}{Coverage-guided test generation for continuous and
  hybrid systems}}.
\newblock {\sl \bibinfo{journal}{Form. Methods Syst. Des.}}
  \bibinfo{volume}{34}(\bibinfo{number}{2}), pp. \bibinfo{pages}{183--213},
  \doi{10.1007/s10703-009-0066-0}.

\bibitemdeclare{inproceedings}{Brim2012}
\bibitem{Brim2012}
\bibinfo{author}{P.~\surnamestart Dluhos\surnameend},
  \bibinfo{author}{L.~\surnamestart Brim\surnameend} \&
  \bibinfo{author}{D.~\surnamestart Safr‡nek\surnameend}
  (\bibinfo{year}{2012}): \emph{\bibinfo{title}{On Expressing and Monitoring
  Oscillatory Dynamics}}.
\newblock In: {\sl \bibinfo{booktitle}{HSB 2012}}, \doi{10.4204/EPTCS.92.6}.

\bibitemdeclare{inproceedings}{Donahue2009}
\bibitem{Donahue2009}
\bibinfo{author}{M.~M. \surnamestart Donahue\surnameend},
  \bibinfo{author}{G.~\surnamestart Buzzard\surnameend} \&
  \bibinfo{author}{A.~E. \surnamestart Rundell\surnameend}
  (\bibinfo{year}{2009}): \emph{\bibinfo{title}{Robust parameter identification
  with adaptive sparse grid-based optimization for nonlinear systems biology
  models}}.
\newblock In: {\sl \bibinfo{booktitle}{ACC Conference}},
  \doi{10.1109/ACC.2009.5160512}.

\bibitemdeclare{inproceedings}{4434348}
\bibitem{4434348}
\bibinfo{author}{R.~\surnamestart Ghaemi\surnameend} \&
  \bibinfo{author}{D.~\surnamestart Del~Vecchio\surnameend}
  (\bibinfo{year}{2007}): \emph{\bibinfo{title}{Evaluating the robustness of a
  biochemical network model}}.
\newblock In: {\sl \bibinfo{booktitle}{Decision and Control, 2007 46th IEEE
  Conference on}}, pp. \bibinfo{pages}{615--620},
  \doi{10.1109/CDC.2007.4434348}.

\bibitemdeclare{article}{DelVecchio2009}
\bibitem{DelVecchio2009}
\bibinfo{author}{R.~\surnamestart Ghaemi\surnameend},
  \bibinfo{author}{J.~\surnamestart Sun\surnameend}, \bibinfo{author}{P.~A.
  \surnamestart Iglesias\surnameend} \& \bibinfo{author}{D.~\surnamestart
  Del~Vecchio\surnameend} (\bibinfo{year}{2009}): \emph{\bibinfo{title}{A
  Method for determining the robustness of bio-molecular oscillator models}}.
\newblock {\sl \bibinfo{journal}{BMC Systems Biology}}
  \bibinfo{volume}{3}(\bibinfo{number}{95}), \doi{10.1186/1752-0509-3-95}.

\bibitemdeclare{article}{Iglesias2006}
\bibitem{Iglesias2006}
\bibinfo{author}{J.~\surnamestart Kim\surnameend}, \bibinfo{author}{D.~G.
  \surnamestart Bates\surnameend}, \bibinfo{author}{I.~\surnamestart
  Postlethwaite\surnameend}, \bibinfo{author}{L.~\surnamestart Ma\surnameend}
  \& \bibinfo{author}{P.~A. \surnamestart Iglesias\surnameend}
  (\bibinfo{year}{2006}): \emph{\bibinfo{title}{Robustness analysis of
  biochemical network models}}.
\newblock {\sl \bibinfo{journal}{IEE Proc. Systems Biology}}
  \bibinfo{volume}{153}(\bibinfo{number}{2}), pp. \bibinfo{pages}{96--104},
  \doi{10.1049/ip-syb:20050024}.

\bibitemdeclare{article}{laub1998molecular}
\bibitem{laub1998molecular}
\bibinfo{author}{M.T. \surnamestart Laub\surnameend} \& \bibinfo{author}{W.F.
  \surnamestart Loomis\surnameend} (\bibinfo{year}{1998}):
  \emph{\bibinfo{title}{A molecular network that produces spontaneous
  oscillations in excitable cells of Dictyostelium}}.
\newblock {\sl \bibinfo{journal}{Molecular biology of the cell}}
  \bibinfo{volume}{9}(\bibinfo{number}{12}), pp. \bibinfo{pages}{3521--3532},
  \doi{10.1091/mbc.9.12.3521}.

\bibitemdeclare{book}{Motwani:1995:RA:211390}
\bibitem{Motwani:1995:RA:211390}
\bibinfo{author}{R.~\surnamestart Motwani\surnameend} \&
  \bibinfo{author}{P.~\surnamestart Raghavan\surnameend}
  (\bibinfo{year}{1995}): \emph{\bibinfo{title}{Randomized algorithms}}.
\newblock \bibinfo{publisher}{Cambridge University Press},
  \bibinfo{address}{New York, NY, USA}, \doi{10.1017/CBO9780511814075}.

\bibitemdeclare{article}{Nonaka20101889}
\bibitem{Nonaka20101889}
\bibinfo{author}{Y.~\surnamestart Nonaka\surnameend},
  \bibinfo{author}{H.~\surnamestart Ono\surnameend},
  \bibinfo{author}{K.~\surnamestart Sadakane\surnameend} \&
  \bibinfo{author}{M.~\surnamestart Yamashita\surnameend}
  (\bibinfo{year}{2010}): \emph{\bibinfo{title}{The hitting and cover times of
  Metropolis walks}}.
\newblock {\sl \bibinfo{journal}{Theoret. Comput. Sci.}}
  \bibinfo{volume}{411}(\bibinfo{number}{16–18}), pp. \bibinfo{pages}{1889 --
  1894}, \doi{10.1016/j.tcs.2010.01.032}.

\bibitemdeclare{book}{Kuznetsov2004}
\bibitem{Kuznetsov2004}
\bibinfo{author}{Kuznetsov \surnamestart Y.\surnameend} (\bibinfo{year}{2004}):
  \emph{\bibinfo{title}{{Elements of Applied Bifurcation Theory }}}.
\newblock \bibinfo{publisher}{Springer}.

\end{thebibliography}

\end{document}